\documentclass[12pt]{article}
\usepackage{amsmath}
\usepackage{amssymb}
\usepackage{graphicx}
\usepackage[retainorgcmds]{IEEEtrantools}
\usepackage{bm}
\usepackage[top=3cm, bottom=3.25cm, left=2.75cm, right=2.75cm]{geometry}

\newcommand{\be}{\begin{equation}}
\newcommand{\ee}{\end{equation}}
\def\beq{\begin{eqnarray}}
\def\eeq{\end{eqnarray}}

\def\ni{\noindent}
\def\bi{\begin{itemize}}
\def\ei{\end{itemize}}

\begin{document} 

\setcounter{page}{0}
\thispagestyle{empty}

\begin{flushright} \small
YITP--SB--11--41  \\
\end{flushright}
\smallskip

\begin{center} \LARGE
\textbf{The Semi-Chiral Quotient, \\   Hyperk\"{a}hler Manifolds and T-duality} \\[6mm] \normalsize

P.  Marcos Crichigno$^{\dagger,}\footnote{crichigno@max2.physics.sunysb.edu} $ 
 \\[3mm]
 {\small\slshape
 $^\dagger$C.N. Yang Institute for Theoretical Physics\\
 State University of New York at Stony Brook, NY 11790, USA\\[3mm]
 }
\end{center}
\vspace{10mm}

\centerline{\bfseries Abstract} \medskip

\ni  We study the construction of generalized K\"{a}hler manifolds, described purely in terms of $\mathcal{N}=(2,2)$ semichiral superfields, by a quotient using the semichiral vector multiplet. Despite the presence of a $b$-field in these models, we show that the quotient of a hyperk\"{a}hler manifold is hyperk\"{a}hler, as in the usual hyperk\"{a}hler quotient.  Thus, quotient manifolds with torsion cannot be constructed by this method. Nonetheless, this method does give a new description of hyperk\"{a}hler manifolds in terms of two-dimensional  $\mathcal{N}=(2,2)$ gauged non-linear sigma models involving semichiral superfields and the semichiral vector multiplet. We give two examples: Eguchi-Hanson and Taub-NUT.  By T-duality, this gives new gauged linear sigma models describing the T-dual of Eguchi-Hanson  and NS5-branes. We also clarify some aspects of T-duality relating these models to $\mathcal{N}=(4,4)$ models for chiral/twisted-chiral fields and comment briefly on more general quotients that can give rise to torsion and give an example.


\small
\vspace{1cm}

\newpage
\tableofcontents{}
\vspace{1cm}
\bigskip\hrule

\normalsize



\section{Introduction}

Recent developments in both physics and mathematics are renewing the interest in general $d=2$, $\mathcal{N}=(2,2)$ sigma models. From the physics perspective, these models describe string compactifications with NS-NS fluxes and, from the mathematics perspective, they provide a useful tool in exploring aspects of Generalized Complex Geometry. This is an example of the interesting interplay between geometry and supersymmetry, initiated by Zumino in the classic work \cite{Zumino:1979et}. It is well known by now that the conditions under which $d=2$, $\mathcal{N}=(1,1)$ sigma models (with no Wess-Zumino term) admit an extended supersymmetry, can be solved by requiring the target space to be K\"{a}hler, for the case of  $\mathcal{N}=(2,2)$, and  hyperk\"{a}hler  for  $\mathcal{N}=(4,4)$ \cite{AlvarezGaume:1981hm}. The action for the sigma model is then simply given by the K\"{a}hler potential $K(\Phi^a, \bar \Phi^a)$ of the target space, with the complex coordinates $\Phi^a$ identified with $\mathcal{N}=(2,2)$ chiral superfields satisfying $\bar{\Bbb D}_+ \Phi^a =\bar{ \Bbb D}_- \Phi^a=0$.  These ideas lead to a variety of  applications of supersymmetric methods to K\"{a}hler geometry. An example of this is the hyperk\"{a}hler quotient \cite{Lindstrom:1983rt,Hitchin:1986ea}. This method is based on the gauging of isometries along directions parametrized by chiral superfields and provides a powerful method for constructing potentials describing hyperk\"{a}hler manifolds. \\

The introduction of a Wess-Zumino term generalizes these models in an interesting way by introducing torsion ($i.e.$, H-flux), leading to what is known as bihermitean geometry \cite{Gates:1984nk}. To describe general bihermitean models in $\mathcal{N}=(2,2)$ superspace, it is necessary to include directions parametrized by twisted-chiral and semichiral superfields. Since the $\mathcal{N}=(2,2)$ vector multiplets introduced in \cite{Lindstrom:2007vc,Lindstrom:2008hx} allow the gauging of isometries in general bihermitean manifolds, it has led us in the present paper to consider more general quotients. This can be used to construct explicit generalized potentials, few of which are known. It would be particularly interesting if one could find potentials describing non-k\"{a}hlerian manifolds (see, $e.g.$, \cite{Kapustin:2006ic} for a discussion of related issues), but it would also be useful to have explicit generalized descriptions of usual hyperk\"{a}hler manifolds. \\

The main goal of this paper is therefore to study certain quotients in a bihermitean setting. We focus on $d=2$ sigma models involving only semichiral superfields, with a $U(1)$ isometry, gauged by the action of the semichiral vector mutiplet. We show that the quotient of a  hyperk\"{a}hler manifold is hyperk\"{a}hler. Thus, despite what could have been naively expected, the resulting quotient manifold has no torsion. We give two explicit examples in four dimensions: Eguchi-Hanson and Taub-NUT. We also perform a T-duality on the latter, which gives us a new $\mathcal{N}=(2,2)$ gauged linear sigma model describing NS5-branes and we briefly discuss a type of quotient that does lead to an H-flux, by incorporating coordinates other than semichiral.\\

The paper is organized as follows. The remainder of this Section contains no new results, but simply reviews some basic elements of general $\mathcal{N}=(2,2)$ models. We focus on the geometry of general semichiral sigma models and the semichiral vector multiplet, which gauges isometries of these sigma models. In Section \ref{The semichiral Quotient} we describe the semichiral quotient and state one of the main results of the paper. In Section \ref{T-Duality} we clarify and extend a duality relation of these semichiral models with $\mathcal{N}=(4,4)$ models for chiral and twisted-chiral fields. In Sections \ref{Eguchi-Hanson} and \ref{Taub-NUT} we give the explicit construction of two well-known gravitational instanton solutions (Eguchi-Hanson and Taub-NUT) and in Section \ref{NS5 Branes} we describe NS5-branes as  a T-dual of Taub-NUT and comment on instanton corrections. In Section \ref{T-dual of Eguchi-Hanson} we present the T-dual of Eguchi-Hanson. We conclude with a summary and discussion of open problems.

\subsection{General $\mathcal{N}=(2,2)$ sigma models}

The models originally studied in  \cite{Zumino:1979et} are not the most general since they don't have a Wess-Zumino term. This motivated the study of general $\mathcal{N}=(1,1)$ models \cite{Gates:1984nk} (see also \cite{Curtright:1984dz,Howe:1985pm}), containing both a metric and a $b$-field, the latter corresponding to a Wess-Zumino term in the action. A general $\mathcal{N}=(1,1)$ sigma model  is described by
\beq
\mathcal{L}=-\frac{1}{4} \int  d^2 \theta ( D_+ \Phi^{\mu})( D_- \Phi^{\nu}) \left( g_{\mu \nu} (\Phi) +b_{\mu \nu} (\Phi)\right),
\label{N=1 Sigma Model}
\eeq
where the $\mathcal{N}=(1,1)$ superfields  $\Phi^{\mu}$ are target space coordinates, $g_{\mu \nu}$ is the target space metric, and  $b_{\mu \nu}=- b_{\nu \mu}$ is the NS-NS 2-form. In the case $b=0$, it reduces to the original case studied by Zumino. 

Studying the conditions under which such models admit an extended supersymmetry led to the discovery  \cite{Gates:1984nk} of a rich geometrical structure: generalized K\"{a}hler geometry.  It was found that, associated to the $\mathcal{N}=(2,2)$ supersymmetry, there are two complex structures $J_{\pm}$ and the metric is hermitean with respect to both. Furthermore, the presence of the $b$-field induces a connection with torsion (proportional to $H=db$) and the complex structures are covariantly constant with respect to this connection. This is what is known as bihermitean geometry. The framework of Generalized Complex Geometry, recently developed by Hitchin \cite{Hitchin:2004ut} and Gualtieri \cite{Gualtieri:2003dx}, describes  this geometry as a generalized K\"{a}hler geometry and we will use these terms interchangeably.

Since the class of models studied by Zumino admits an explicit $\mathcal{N}=(2,2)$ formulation, it is natural to wonder if these  general models admit such a description and, indeed, they do. This was made possible by the introduction of  $\mathcal{N}=(2,2)$ twisted-chiral superfields satisfying $\bar{\Bbb D}_+ \chi = \Bbb D_- \chi=0$.  
 One considers a scalar function, depending both on chiral and twisted-chiral superfields, $i.e.$,  $K(\Phi, \bar \Phi, \chi, \bar \chi)$, as a potential for the bihermitean geometry. Due to the twisted nature of the constraints on $\chi$ (relative to $\Phi$), upon reduction to $\mathcal{N}=(1,1)$, one finds an action of the type (\ref{N=1 Sigma Model}) with a non-zero $b_{\mu \nu}$, provided by cross-terms like $K_{\Phi \bar \chi}$.  Interestingly, these models fall outside the classification of \cite{AlvarezGaume:1981hm}. Indeed,  when the condition $K_{\bar \Phi \Phi}+K_{\bar \chi \chi}=0$ is satisfied, the model has $\mathcal{N}=(4,4)$ supersymmetry without being hyperk\"{a}hler \cite{Gates:1984nk}. An example of this is the $S^3 \times S^1$ WZW model \cite{Rocek:1991vk}.

For some time, however, it remained unclear what set of $\mathcal{N}=(2,2)$ fields provides a \textit{complete} description of bihermitean geometry. In $d=2$, $\mathcal{N}=(2,2)$ superspace has four fermionic coordinates $\theta^{\pm}, \bar{\theta}^{\pm}$, where the $\pm$ index stands for chirality under Lorentz transformations, and $ \bar{\theta}^{\pm}= (\theta^{\pm})^{\ast}$. Thus, the most general, linear,  SUSY-invariant, constraints one can impose are \cite{Sevrin:1996jr}
\beq  \nonumber
\bar{\Bbb D}_+ \Phi = \bar{ \Bbb D}_- \Phi=0 && \text{Chiral} \\ \label{N=2 fields}
\bar{\Bbb D}_+ \chi = \Bbb D_- \chi=0  && \text{Twisted chiral} \\ \nonumber
\bar{\Bbb D}_+ \Bbb X_L= \bar{\Bbb D}_- \Bbb X_R=0  &&  \text{Left and Right semichiral} .  \nonumber
\eeq
It was believed at the time, and explicitly conjectured in  \cite{Sevrin:1996jr}, that this is the set of fields\footnote{To be able to integrate out the auxiliary $\mathcal{N}=(1,1)$ spinor superfields, it is necessary to have the same number of left and right  semichiral fields.} which gives the most general description of  an $\mathcal{N}=(2,2)$ sigma model. This was finally proven\footnote{To avoid the reader's confusion, it is worth mentioning that the conclusion in  Ref. \cite{Bogaerts:1999jc} (which includes other important results), that this is not the case, is erroneous. See  \cite{Lindstrom:2005zr} for an explanation.} in \cite{Lindstrom:2005zr}. The action is given in terms of a generalized K\"{a}hler potential
\beq
\mathcal{L}=\int d^{2} \theta d^{2} \bar \theta \, K(\Phi, \bar \Phi, \chi , \bar \chi,\Bbb X_L, \Bbb{ \bar X}_L ,\Bbb X_R,\Bbb{ \bar X}_R)\,.
\eeq
From the constraints (\ref{N=2 fields}) one sees that the generalized potential $K$ is defined up to  generalized K\"{a}hler transformations $f(\phi,\chi,\Bbb X_L)+g(\phi,\bar \chi, \Bbb X_R )+\bar f(\bar \phi, \bar \chi, \bar{ \Bbb X}_L)+\bar g(\bar \phi, \chi, \bar{ \Bbb X}_R )$, since these vanish upon integration in superspace. Upon reduction to $\mathcal{N}=(1,1)$ fields, the action has the form (\ref{N=1 Sigma Model}) and one can read off the metric and $b$-field completely in terms of second derivatives of $K$. \\

A comment on notation: $\mathcal{N}=(2,2)$ spinor derivatives are denoted by $\Bbb{D}_{\pm}$ to distinguish them from the $\mathcal{N}=(1,1)$ derivatives $D_{\pm}$. We usually denote the lowest $\mathcal{N}=(1,1)$ components of chiral and twisted-chiral fields by the same letters as the $\mathcal{N}=(2,2)$ fields, whereas for semichiral fields we write $\Bbb X_{L,R}|=X_{L,R}$. When writing the metric and $b$-field, it should be understood that we are referring to the $\mathcal{N}=(1,1)$ components.

\subsection{Geometry of semichiral sigma models}

Consider a non-linear sigma model of a set of semichiral superfields $\Bbb X_L^a, \Bbb X_R^{a'}$, $a,a'=1,...,d_s$ with an action given by 
\beq
\mathcal{L}=\int{d^{2} \theta d^{2} \bar \theta K \left( \Bbb  X_L, \Bbb{ \bar X}_L, \Bbb X_R,   \Bbb{ \bar X}_R\right)}\,.
\eeq
These models were first studied in \cite{Buscher:1987uw}, showing that upon reduction to $\mathcal{N}=(1,1)$, they lead to a general non-linear sigma model of the type of (\ref{N=1 Sigma Model}). However, semichiral superfields are less constrained  than chiral and twisted-chiral fields and contain auxiliary superfields which, when integrated out, induce non-linearities in the $\mathcal{N}=(1,1)$ action. As a consequence, the metric and $b$-field are non-linear functions of second derivatives of $K$. These can be written compactly \cite{Bogaerts:1999jc,Lindstrom:2005zr} in terms of the complex structures $J_{\pm}$ and a closed 2-form $\Omega$ as 
\beq \label{G and B}
g=\Omega\, [J_+,J_-], && b=\Omega \, \{J_+,J_-\}\,.
\eeq
The complex structures and $\Omega$ are completely determined by the generalized potential by
\beq \label{Complex Structures}
J_+=\left( \begin{array}{cc} J_s &0 \\ \mathcal{K}_{RL}^{-1} C_{LL} & \mathcal{K}_{RL}^{-1}J_s \mathcal{K}_{LR} \end{array} \right), && J_-=\left( \begin{array}{cc} \mathcal{K}_{LR}^{-1}J_s \mathcal{K}_{RL} & \mathcal{K}_{LR}^{-1} C_{RR} \\ 0 & J_s \end{array} \right)\,, 
\eeq
where $J_s$ is a $2d_s$-dimensional matrix of the form $\text{diag}(i,-i)$ and 
\beq
\Omega=\left( \begin{array}{cc} 
 0 & \mathcal{K}_{LR} \\ 
-(\mathcal{K}_{LR})^t &0
 \end{array} \right)\,,
\eeq
with
 \beq \label{K}
 \mathcal{K}_{LL}=\left( \begin{array}{cc} K_{LL} &K_{L \bar L} \\ K_{\bar L L} & K_{\bar L \bar L} \end{array} \right), &&\mathcal{K}_{LR}=\left( \begin{array}{cc} K_{L R} &K_{L \bar R} \\ K_{\bar L R} & K_{\bar L \bar R} \end{array} \right)\,, 
 \eeq
where $K_{LR}\equiv \frac{\partial^2 K}{\partial X_L \partial X_R}$, etc. and $\mathcal{K}_{LR}^{-1}\equiv(\mathcal{K}_{RL})^{-1}$. \\

In four dimensions ($i.e.$, $d_s=1$) there's an additional structure, leading to the anti-commutator of the complex structures to being proportional to the identity, namely
\beq  \label{Anticommutator Identity0}
\{J_+,J_-\}=c \,\Bbb I,
\eeq
where $c$ is a scalar function given by 
\beq \label{c in 4 dimensions}
 c=-2 \frac{|K_{LR}|^2+|K_{L \bar R}|^2-2 K_{L \bar L} K_{R \bar R}}{|K_{LR}|^2-|K_{L \bar R}|^2} \, .
 \label{c}
\eeq
As we shall review next, it contains important information about the geometry; when $c$ is a constant and $|c|<2$, the manifold is hyperk\"{a}hler.

\subsection{Hyperk\"{a}hler case} 
\label{hyper-Kahler case}

As shown in  \cite{Lindstrom:2005zr}, a generalized K\"{a}hler manifold of $4N$ real dimesions, described in terms of semichiral superfields,  is hyperk\"{a}hler if $\{J_+,J_-\}=c \, \Bbb I$ with $c$ a constant and $|c|<2$ (see also \cite{Bogaerts:1999jc} for the particular case $c=0$).  This is easy to see from the expression for the $b$-field in  (\ref{G and B}); since $\Omega$ is a closed 2-form, the torsion, $H = d b= \Omega \, dc$, vanishes for constant $c$. If the manifold is hyperk\"{a}hler, there must be three complex structures and, indeed, a third complex structure $J_3$ can be constructed from $J_{\pm}$ by
\beq
J_3=\frac{1}{\sqrt{\left(\frac{2}{c}\right)^2-1}}\left(\Bbb I- \frac{2}{c} J_+J_- \right)\, .
\eeq
A trivial example of a hyperk\"{a}hler manifold (and one which will be used in what follows) is flat $\Bbb R^{4n}$ with a constant $b$-field. This is described by the generalized potential
\beq
K_{\Bbb R^{4n}}=   \sum_{i=1}^n \left( \Bbb{\bar{X}}_L^i  \Bbb  X_L^i +\Bbb{\bar{X}}_R^i  \Bbb X_R^i +\alpha ( \Bbb{\bar{X}}_R^i \Bbb  X_L^i+\Bbb{ \bar{X}}_L^i \Bbb  X_R^i  )\right)\, .
\eeq
From equations (\ref{G and B}-\ref{K}), one finds the (constant) metric, $b$-field, and complex structures satisfying 
\beq
\{J_+,J_-\}=2(1-\frac{2}{\alpha ^2})\Bbb I \, .
\eeq
For the metric to be positive definite, $\alpha^2>1$ is required, which also ensures $|c|<2$. For the special value $\alpha^2 =2$, the $b$-field vanishes.

\subsection{Semichiral vector multiplet}

The semichiral vector multiplet \cite{Lindstrom:2007vc, Lindstrom:2008hx} was introduced to gauge isometries along semichiral directions, $e.g.$,
\beq
\delta \Bbb X_L = i \lambda, &&  \delta \Bbb X_R = i \lambda \,.
\eeq
It is described in terms of three real supervector fields $V^{\alpha}=(V_L,V_R,V')$, with gauge transformations 
\beq
\delta  V_L = i( \bar \Lambda_L -  \Lambda_L)\, , \qquad \delta  V_R = i(\bar  \Lambda_R -  \Lambda_R)\, , \qquad \delta  V' =(\Lambda_R + \bar  \Lambda_R- \Lambda_L -\bar  \Lambda_L)\, . 
\eeq
It's convenient to introduce the complex combinations
\beq
\Bbb V=\frac{1}{2}(-V'+i (V_L-V_R))\,, \qquad \tilde{ \Bbb V} =\frac{1}{2}(-V'+i (V_L+V_R))\,,
\eeq
with gauge transformations
\beq
\delta \Bbb V = \Lambda_L-\Lambda_R\,, && \delta   \tilde{ \Bbb V} =  \Lambda_L-\bar \Lambda_R\,.
\eeq
The corresponding chiral and twisted-chiral field strengths are
\beq
\Bbb  F=\bar{ \Bbb D}_+ \bar{\Bbb D}_- \Bbb  V\,, \qquad \tilde{ \Bbb  F} = \bar{\Bbb D}_+ \Bbb D_- \tilde{ \Bbb V}\,.
\eeq
Thus, the nonvanishing commutation relations are \cite{Lindstrom:2008hx} 
\beq \nonumber
 \{\bar \nabla_{\pm},  \nabla_{\pm}\}= i \mathcal{D}_{\pm \pm}\,, \qquad  i \{\bar \nabla_+, \bar \nabla_-\}= \Bbb F\,, \qquad i \{\bar \nabla_+, \nabla_-\} = \tilde{\Bbb F }\, ,
\eeq
where $\nabla_{\pm}$ are gauge-covariant superderivatives. The kinetic terms for the gauge fields are given by
\beq \label{kinetic terms}
\mathcal{L}_{gauge}=  \int d^4 \theta \frac{1}{e^2} ( \Bbb F  \bar{\Bbb F}  -\Bbb{ \tilde F}  \Bbb{ \bar{\tilde F}}  )\, .
\eeq
It's also possible to add Fayet-Iliopoulos (FI) terms of the form
\beq \label{FI terms}
\mathcal{L}_{FI}=-\int d^4 \theta \,(t \Bbb V +s \tilde{ \Bbb V} +c.c.)= -\int d^4 \theta \, t_{\alpha} V^{\alpha} \, ,
\eeq
where we defined $t_{\alpha}\equiv- \left(\text{Im}(s+t),\text{Im}(s-t),\text{Re}(s+t)\right)$. These will play an important role in what follows. Upon reduction to $\mathcal{N}=(1,1)$,  (\ref{kinetic terms}) gives  the usual kinetic terms. The only dimensionful scale is $[e]=1$  and the low energy limit corresponds to taking $e \to \infty$. Therefore, the kinetic terms are irrelevant in the IR limit and the gauge fields $V',V_L,V_R$ become non-dynamical and are integrated out. 
Thus, the gauged linear sigma model will flow in the IR to a non-linear sigma model given by a semichiral quotient, which we now describe.

\section{The Semichiral Quotient}
\label{The semichiral Quotient}

Here we describe what we refer to as the semichiral quotient. We consider a bihermitean manifold $\mathcal{M}$ of $d=4(N+1)$ real dimensions, parameterized by semichiral coordinates $(\Bbb  X^{a}_{L},\Bbb X^{a'}_R)$ with $a,a'=1,...,N+1$ and generalized potential $K(\Bbb X^{a}_{L},\Bbb X^{a'}_R)$. We assume the existence of a $U(1)$ Killing vector
\beq
k=k^{a}\partial_{a}+k^{\bar a}\partial_{\bar a}+k^{a'}\partial_{a'}+k^{\bar a'}\partial_{\bar a'}\,,
\eeq
generating the isometry 
\beq
\delta X = \left[ \lambda k ,X\right]\,,
\eeq
where $\lambda$ is the parameter of the transformation and $X$  is any of the coordinates.  We now choose coordinates $(\Bbb X^a_L,\Bbb X^{a'}_R)=(\Bbb X^{i}_{L},\Bbb X^{i'}_R; \Bbb X_L,\Bbb X_R)$, with $i,i'=1,...,N$, which are adapted to the isometry and the Killing vector takes the form
\beq
k=i\left( \partial_L -\bar{\partial}_L +\partial_R -\bar{\partial}_R\right)\,.
\eeq
In these adapted coordinates, the generalized potential depends explicitly on the $4N$ neutral coordinates $(\Bbb X^{i}_{L},\Bbb X^{i'}_R)$ and the 3 invariant combinations $\Bbb X^{\alpha}=(\Bbb X_L+\bar{\Bbb  X}_L,\Bbb X_R+ \bar{\Bbb  X}_R,i (\Bbb X_R - \bar{\Bbb  X}_R  -\Bbb  X_L +\bar{\Bbb  X}_L))$. Now we proceed to  gauge this isometry by promoting the parameter $\lambda$ to a corresponding semichiral field and introducing a semichiral vector multiplet. Then, the function $\hat K$ is defined by 
\beq
\hat K(\Bbb X^{i}_{L},\Bbb X^{i'}_R)&=& K(\Bbb X^{i}_{L},\Bbb X^{i'}_R;\Bbb X^{\alpha}+V^{\alpha}) - t_{\alpha} V^{\alpha}\, , 
\eeq
where $V^{\alpha}=V^{\alpha}(\Bbb X^i_{L,R})$ is given by solving its equations of motion
\beq
\frac{\partial  K(\Bbb X^{i}_{L},\Bbb X^{i'}_R;\Bbb X^{\alpha}+V^{\alpha})}{ \partial V^{\alpha}}&=&t_{\alpha}\, 
\eeq
and choosing the gauge $X^{\alpha}=0$. The new potential $\hat K$ depends on $4N$ coordinates (and three FI parameters $t_{\alpha}$), and describes the quotient manifold $\hat{ \mathcal{M}}$ of real dimension $4N$.

Now we state one of our main results. Assume that $\mathcal{M}$ is a hyperk\"{a}hler manifold and therefore
\beq  \label{Anticommutator Identity}
\{J_+,J_-\}=c \,\Bbb I\, ,
\eeq
with $c$ a constant, as discussed in Section \ref{hyper-Kahler case}. Then, the anticommutator of the complex structures on the quotient manifold $\mathcal{\hat M}$  is given by
\beq \label{c Hat}
\{\hat{J}_+,\hat{J}_-\}=c\, \Bbb I\, 
\eeq
(with the same $c$ on the right-hand side). In particular, this implies that the quotient manifold is also hyperk\"{a}hler. In the current setting, the proof of (\ref{c Hat}) requires some rather tedious algebra, but is straightforward. Imposing  (\ref{Anticommutator Identity}) leads to the set of equations
\beq
\left\{\mathcal{K}_{LR}^{-1}C_{RR}\mathcal{K}_{RL}^{-1},J_s\right\}&=&0\, ,  \label{off-diagonal} \\[2mm] 
J_s \mathcal{K}_{LR}^{-1}J_s \mathcal{K}_{RL}  +   \mathcal{K}_{LR}^{-1} J_s \mathcal{K}_{RL}J_s +\mathcal{K}_{LR}^{-1} C_{RR} \mathcal{K}_{RL}^{-1} C_{LL}&=& c\,  \Bbb{I}\, , \label{diagonal} 
\eeq
and those which follow from these exchanging ($L\leftrightarrow R$). Using standard relations between second derivatives of Legendre-transformed functions, and identities for matrix inverses, we show that these equations also hold for $\hat K$, proving the assertion (\ref{c Hat}) (see Appendix \ref{semichiral Quotient} for more details). 

A brief comment is in order. In showing that the structure (\ref{Anticommutator Identity}) is preserved by the quotient, we have actually not made use of the fact that $c$ is a constant. Thus, one could in principle extended our results to bihermitean geometries satisfying (\ref{Anticommutator Identity}), other than hyperk\"{a}hler (with $c$ an arbitrary function), if there are any such manifolds. This, however, is not the case due to the following result \cite{HitchinRocek}. Although the set of equations (\ref{off-diagonal}, \ref{diagonal}) are satisfied identically in four dimensions, they highly restrict the geometry in higher dimensions. So much indeed, that the only manifolds satisfying (\ref{Anticommutator Identity}) in $d \geq 8$ are those with a constant $c$, $i.e.$, hyperk\"{a}hler manifolds.  

\subsection{Geometrical interpretation}
\label{Geometrical interpretation}
It might seem surprising at first that the semichiral quotient coincides with the hyperk\"{a}hler quotient. However, this is clarified by the following geometrical interpretation \cite{BigPaper}. The hyperk\"{a}hler quotient \cite{Lindstrom:1983rt,Hitchin:1986ea} is based on assuming the existence of three symplectic 2-forms $\omega^p$, $p=1,2,3$, and a triholomorphic Killing vector $k$, $i.e.$,
\beq
\mathcal{L}_{k} \omega^p = i_{k} d \omega^p+d(i_k \omega^p)= 0\,.
\eeq
Since $d\omega^p=0$, this implies the existence (locally) of the three moment maps, $\mu^p$, such that
\beq
i_k \omega^p= d \mu^p\,.
\eeq
Setting the moment maps to zero (and dividing by the isometry), leads to the hyperk\"{a}hler quotient. The relation with the semichiral quotient is based on the observation that if $\left[ J_+,J_-\right]$ is invertible (which requires the presence of only semichiral fields), the closed 2-form 
\beq
\Omega=g \left[J_+,J_{-} \right]^{-1}
\eeq
is well defined\footnote{Even in the presence of only semichiral fields, $\left[ J_+,J_-\right]$ can fail to be invertible at some points or loci in the manifold, leading to type change. We shall not consider this case here.}. This symplectic form can be decomposed \cite{Bogaerts:1999jc} into its holomorphic and anti-holomorphic part, with respect to both complex structures $J_{\pm}$, $i.e.$,
\beq\label{reality condition}
\Omega=\Omega^{(2,0)}_- + \bar \Omega^{(0,2)}_- = \Omega^{(2,0)}_+ + \bar \Omega^{(0,2)}_+\,
\eeq
and $d \Omega=0$ implies
\beq \label{decomposition omega}
 \partial \Omega^{(2,0)}_{\pm}=\bar \partial \Omega^{(2,0)}_{\pm}=0\,,
 \eeq
and the complex conjugates. This implies the existence of four moment maps $\mu_{\pm}, \bar \mu_{\pm}$, subject to the reality condition 
\beq
\mu_-+\bar \mu_- = \mu_+ +\bar \mu_+\,,
\eeq
which follows from (\ref{reality condition}). Thus, there are three independent moment maps and the semichiral quotient coincides with the hyperk\"{a}hler quotient. \\[2mm]

It can also be understood \cite{HitchinRocek} in these geometrical terms why only hyperk\"{a}hler manifolds satisfy (\ref{Anticommutator Identity}). In a generalized K\"{a}hler manifold, the 3-form $H=db$ has no $(3,0)$ or $(0,3)$ part (see, $e.g.$, \cite{Gualtieri:2003dx}) with respect to both $J_{\pm}$, $i.e.$,
\beq
H=  H^{(1,2)}_{\pm}+H^{(2,1)}_{\pm}\,.
\eeq
Assuming (\ref{Anticommutator Identity}), one has $H=\Omega \, dc$. Using (\ref{reality condition}) and $dc =\partial c +\bar \partial c$, one sees that $(3,0)$ and $(0,3)$ parts appear. The requirement that they vanish implies
\beq
\partial c = \bar \partial c=0\,.
\eeq
Thus, $c$ is a constant and $H$ vanishes completely. 

\subsection{Comment on more General Quotients}
\label{Comment on general Quotients}

As we have just seen, quotients involving only semichiral fields will not lead to a non-trivial $b$-field. However, considering several types of fields typically does. Here we give a simple example. Consider a set of semichiral fields and a single chiral field $\Phi$, gauged by the usual vector multiplet $V$, $i.e.$,
\beq
K=\Bbb{\bar{X}}_L e^V  \Bbb  X_L +\Bbb{\bar{X}}_R e^V \Bbb X_R +\alpha \left( \Bbb{\bar{X}}_R e^V \Bbb  X_L+\Bbb{ \bar{X}}_L e^V \Bbb  X_R \right)+t \bar \Phi e^V \Phi- r V\,.
\eeq
Integrating out $V$ (and choosing the gauge $\Phi=1$) leads to 
\beq  \label{K quotient semi and chiral}
K=r \log{\left(\Bbb{\bar{X}}_L  \Bbb  X_L +\Bbb{\bar{X}}_R \Bbb X_R +\alpha \left( \Bbb{\bar{X}}_R \Bbb  X_L+\Bbb{ \bar{X}}_L \Bbb  X_R \right)+t \right)}\,.
\eeq
From (\ref{c in 4 dimensions}) we find
\beq
c=-2 + \frac{4 t \left(\alpha ^2-1\right)}{\alpha(\alpha t-R)}\,,
\eeq
where $R\equiv \bar X_R X_L + \bar X_L X_R + \alpha (\bar X_L X_L +\bar X_R X_R)$. Thus $c$ is not a constant and there's a non-trivial $b$-field. Although we will not analyze this model in full detail here, we can already study some features.  From (\ref{K quotient semi and chiral}), one sees that the limit $t\rightarrow \infty$ corresponds to flat space, while  $t\rightarrow 0$ gives a singular metric. For finite $t$, the metric becomes singular ($c=\pm2$) for $R=t/\alpha$ and $R\rightarrow \infty$.

\section{T-Duality}
\label{T-Duality}

A duality relation between hyperk\"{a}hler manifolds, described in terms of semichiral superfields with $c=0$, and $\mathcal{N}=(4,4)$ models for chiral/twisted-chiral fields was described in  \cite{Bogaerts:1999jc}. Actually, understanding this relation was  one of the motivations for introducing the new vector multiplets and studying T-duality  \cite{Lindstrom:2007sq}. In this Section we would like clarify the exact relation of the duality in  \cite{Bogaerts:1999jc} to T-duality and offer a geometrical interpretation, which also allows us to consider K\"{a}hler manifolds with $c  \neq 0$  and even non-K\"{a}hlerian manifolds (that may still have $\mathcal{N}=(4,4)$). As we shall see, this depends on the character of the isometry along which the duality is performed. We first discuss T-duality along a translational isometry, which leads to a hyperk\"{a}hler manifold. Then, we discuss T-duality along a general isometry.
 
 \subsection{Translational isometry}
 
The duality described in  \cite{Bogaerts:1999jc} involves two steps.  Given a potential $ \hat F(\Phi, \bar \Phi, \chi, \bar \chi)$ satisfying the Laplace equation, one first constructs a potential $ F(\Phi, \bar \Phi, \chi, \bar \chi)$. Then, one performs a Legendre transformation to semichiral superfields. It is the first step which we reinterpret as a rotation of the $\mathcal{N}=(1,1)$ components by a fixed angle. As we shall see below, considering an arbitrary\footnote{The author wishes to thank Martin Rocek for this suggestion.} (constant) rotation by an angle $\nu$ leads to a non-zero (constant) $c$.\\

Consider a potential $\hat F(\phi, \bar \phi, \chi, \bar \chi)$ and assume that there's a translational isometry, generated by the Killing vector
\beq \label{Translational Killing Vector}
k=i(\partial_{\phi}-\partial_{\bar \phi}-\partial_{\chi} +\partial_{\bar \chi})\, .
\eeq 
Thus, in adapted coordinates
\beq
\hat F = \hat F\left(\phi+\bar \phi,\chi+\bar \chi, i (\phi-\bar \phi +\chi -\bar \chi)\right)\, .
\eeq
Assume now that the potential describes an $\mathcal{N}=(4,4)$ model and, therefore, satisfies the Laplace equation
\beq
\hat F_{\phi \bar \phi}+ \hat F_{\chi \bar \chi}=0\, .
\eeq
The important observation now is that a rotation among the $\mathcal{N}=(1,1)$ fields, $(\phi, \chi)$, is allowed and preserves the Laplace equation. Then, when integrating up to the $\mathcal{N}=(2,2)$ potential, one must choose what to call a chiral or twisted-chiral field and we choose to take the rotated fields. That is, we consider the transformation
\beq
\phi \rightarrow  \cos({\nu}) \phi + \sin({\nu}) \chi \, , && \chi \rightarrow \cos({\nu}) \chi -\sin({\nu}) \phi\, .
\eeq
For convenience, we introduce $\theta = \nu + \frac{\pi}{4}$ and \textit{define} the  potential $F(\Phi, \bar \Phi, \chi, \bar \chi)$ by
\beq
F= \hat F\left(\Phi+\bar \Phi, \chi+\bar \chi, i(c(\Phi-\bar \Phi) +s(\chi -\bar \chi))\right)\, ,
\eeq
where we have abbreviated $\cos(\theta)=c, \sin(\theta)=s$. The Killing vector is now  given by
\beq \label{rotated Killing vector}
k=i[(s(\partial_{\Phi}-\partial_{\bar \Phi})-c(\partial_{\chi} -\partial_{\bar \chi})]\,,
\eeq
which implies the transformations for the matter fields
 \beq
\delta \Phi = i s \lambda \, , && \delta \chi = -i c  \lambda\, .
\eeq
This isometry can be gauged by the Large Vector Multiplet (LVM) \cite{Lindstrom:2007vc,Lindstrom:2008hx}, defined similarly to the SVM by
\beq
\Bbb V_L = \frac{1}{2} \left(-V'+ i ( V^{\phi}-V^{\chi})\right)\, , && \Bbb V_R = \frac{1}{2} \left(-V'+ i ( V^{\phi}+ V^{\chi})\right)\,, 
\eeq 
where the real vector fields $V^{\alpha}=(V^{\phi}, V^{\chi},V')$ transform as
\beq
\delta V^{\phi}=i(\bar \Lambda - \Lambda)\,, \qquad \delta V^{\chi}=i(\bar{\tilde \Lambda} - \tilde \Lambda)\,,  \qquad \delta V'=-( \Lambda +\bar \Lambda) + \tilde \Lambda+\bar{\tilde \Lambda}\,.
\eeq
Following \cite{Lindstrom:2007sq}, we perform a T-duality to  semichiral fields by defining
\beq \nonumber \label{LVM gauging}
K(\Bbb X_L,\Bbb X_R)&=& F\left(\Phi + \bar \Phi +s V^{\phi},\chi +\bar \chi + c V^{\chi},  i(c(\Phi-\bar \Phi) +s(\chi -\bar \chi)) - csV'\right) \\ &&-\left[ \Bbb X_L \Bbb V_L +\Bbb X_R \Bbb V_R +c.c. \right]\,.
\eeq
In the gauge $\Phi=\chi=0$, we have
\beq \label{T-dual Sevrin} \nonumber
K(\Bbb X_L,\Bbb X_R)&=& F(s V^{\phi}, c V^{\chi}, -csV')-\frac{1}{2} \Big[ iV^{\phi}(\Bbb X_L -\bar{\Bbb X}_L +\Bbb X_R -\bar{\Bbb  X}_R)  \\
&&- i V^{\chi} (\Bbb X_L -\bar{\Bbb X}_L -\Bbb X_R +\bar{\Bbb  X}_R )-  V' (\Bbb X_L +\bar{\Bbb X}_L +\Bbb X_R +\bar{\Bbb  X}_R ) \Big]\,.
\eeq
Integrating out the LVM, $i.e.$, solving
\beq
\frac{\partial K}{\partial V^{\alpha}}=0
\eeq
for the vector fields $V^{\alpha}$ leads to the semichiral potential. From the definition (\ref{c in 4 dimensions}), and using standard implicit differentiation relations (see Appendix \ref{T-duality appendix} for more details), we find a non-zero $c$ given by
\beq \label{c translational Killing vector}
c=-2 \cos(2 \theta)\, . 
\eeq
For the particular case $\theta = \pi/4$, this reduces to the duality described in   \cite{Bogaerts:1999jc}. \\

A short observation that will be useful later\footnote{In coming Sections we will perform T-duality transformations in the other direction, namely from semichiral fields to chiral/twisted-chiral by the use of the semichiral vector multiplet. We expect, however, the same relations to hold.} is that one may alternatively rescale the fields $\phi,\chi$ in (\ref{rotated Killing vector}) to bring the Killing vector to its usual form. Then, the potential $F$ will satisfy a \textit{scaled} Laplace equation:  If $K$ describes a hyperk\"{a}hler manifold with a constant $c= 2(1-\frac{2}{\alpha ^2})$, the dual potential satisfies
\beq \label{Scaled Laplace Equation}
F_{\phi \bar \phi}+(\alpha^2-1)F_{\chi \bar \chi}=0\, .
\eeq

\subsection{General isometry}

As we have just discussed, T-dualizing an $\mathcal{N}=(4,4)$ model along a translational isometry using the LVM leads to a hyperk\"{a}hler manifold, described in terms of semichiral fields. In showing this, the form of the Killing vector was crucial. Indeed, if it acts by translation on $\Phi$ and $\chi$ by \textit{equal} amounts, then $c=0$, while if it acts by different amounts, it leads to a non-zero (but constant) $c$. We wish to investigate now what happens for a general isometry of the form $k=k^{\Phi}(\Phi)\partial_{\Phi} + k^{\chi}(\chi)\partial_{\chi}+ c.c.\,$. If $K$ is invariant under the isometry, the gauging along a general Killing vector is given by \cite{Lindstrom:2007sq}
\beq
K^{(g)}=\exp \left( -\frac{1}{4} V^{\phi} \mathcal{L}_{(J_++J_-)k}-\frac{1}{4} V^{\chi} \mathcal{L}_{(J_+-J_-)k} -\frac{1}{4} V^{'} \mathcal{L}_{J_+J_-k} \right) K\, .
\eeq 
By implicit differentiation (again, see Appendix \ref{T-duality appendix} for details), we find
\beq \label{c General Killing vector }
c=2\left. \left(\frac{|k^{\Phi}|^2-|k^{\chi}|^2}{|k^{\Phi}|^2+|k^{\chi}|^2}\right) \right| _{ \partial K/ \partial V=0}\, .
\eeq
Note that although this expression does not depend on the potential explicitly, it does depend on it implicitly; to write the right-hand side in terms of semichiral coordinates, the relation of chiral/twisted-chiral fields to semichiral fields given by the Legendre transform is needed. In the case $k^{\Phi}=- k^{\bar \Phi}=i \cos(\theta)$ and $k^{\chi}=-k^{\bar \chi}=i \sin(\theta)$, we recover (\ref{c translational Killing vector}). We conclude from (\ref{c General Killing vector }) that for a general isometry $c$ will not be a constant and the dual geometry will not be hyperk\"{a}hler, even if the T-duality preserves the supersymmetry (the isometries preserving $\mathcal{N}=(4,4)$ in this context are translational and rescaling \cite{N=4 and t-duality}).

As an example, consider the gauging of the isometry along the $S^1$ in the $SU(2)\times U(1)$ WZW model, described in terms of chiral/twisted-chiral superfields  \cite{Rocek:1991vk}, recently studied in \cite{Sevrin:2011mc}. The isometry in this case acts by a rescaling of the fields, $i.e.$, 
\beq \label{Rescaling Killing Vector}
k=\Phi \partial_{\Phi}+\bar \Phi \partial_{\bar \Phi}+\chi \partial_{\chi} +\bar \chi \partial_{\bar \chi}\,.
\eeq 
T-dualizing along this direction, the dual potential again describes an $SU(2)\times U(1)$ WZW model, which is not hyperk\"{a}hler. Indeed,  from (\ref{c General Killing vector }), one finds
\beq
c=\frac{2}{\sqrt{1-4 e^{-X'}}}\,,
\eeq
where $X'=X_L+\bar X_L+X_R+\bar X_R$. Since the isometry in the $SU(2)\times U(1)$ WZW model corresponds to a rescaling, the semi-chiral description of this space allows for $\mathcal{N}=(4,4)$.

\section{Eguchi-Hanson}
\label{Eguchi-Hanson}

Here we give the first example of the semichiral quotient. We consider $\Bbb R^8=\Bbb R^4 \times \Bbb R^4$,  described by two copies of a left and right semichiral field, $(\Bbb X_L^{(1)}, \Bbb X_R^{(1)})$ and  $(\Bbb X_L^{(2)},\Bbb X_R^{(2)})$, as discussed in Section \ref{hyper-Kahler case}. We assign equal\footnote{It is worth mentioning that one can invert the charge of one of the pairs, say  $(\Bbb X_L^{(2)},\Bbb X_R^{(2)})$, by dualizing to fields  $\tilde{\Bbb X}_L,\tilde{\Bbb X}_R$ that impose the semichiral constraints on the original pair \cite{hep-th/9801080}. This duality is not based on an isometry and does not change the geometry. Hence, we expect the quotient involving two pairs of semis, either with charges $(+,+)$ or $(+,-)$, to lead to the same geometry.} $U(1)$ charges $q_1=q_2=1$ to both and proceed as described, defining
\beq \label{Gauge action for EH} 
 \hat K&=&  \sum_{i=1,2} \left[  \Bbb{\bar{X}}_L^{(i)} e^{V_L} \Bbb  X_L^{(i)} +\Bbb{\bar{X}}_R^{(i)} e^{V_R}  \Bbb X_R^{(i)} +\alpha ( \Bbb{\bar{X}}_R^{(i)} e^{-i \tilde{\Bbb  V}} \Bbb  X_L^{(i)}+\Bbb{ \bar{X}}_L^{(i)} e^{i \bar{\tilde{ \Bbb V}}} \Bbb  X_R^{(i)})\right] - t_{\alpha}  V^{\alpha}\,.
\eeq
Based on our results of Section \ref{The semichiral Quotient}, we know the resulting quotient manifold will be hyperk\"{a}hler, with $c= 2(1-\frac{2}{\alpha ^2})$. We show below that this is actually the well-known Eguchi-Hanson manifold. Before showing this explicitly, by computation of the quotient potential and metric, we show that this quotient construction actually reduces to the usual hyperk\"{a}hler quotient construction of Eguchi-Hanson in terms of $\mathcal{N}=1$ fields.

\subsection{Reduction to $\mathcal{N}=(1,1)$: Comparison to the hyperk\"{a}hler quotient}
\label{reduction to N=1}

The procedure to reduce to  $\mathcal{N}=(1,1)$ is well known (see, $e.g.$, \cite{Lindstrom:2005zr} for a review). One decomposes the $\mathcal{N}=(2,2)$ gauge-covariant (super)derivatives into their real and imaginary part, namely
\beq
\nabla_{\pm}=\frac{1}{2}\left(  \mathcal{D}_{\pm} -i Q_{\pm}\right), && \bar{\nabla}_{\pm}=\frac{1}{2}\left(  \mathcal{D}_{\pm} + i Q_{\pm}\right)\,. 
\eeq
We perform the reduction of the matter fields $\Bbb X_L, \Bbb X_R$ in the covariant approach (see Appendix \ref{AppendixN=1} for more details), defining
\beq
\bar{\Hat{ \Bbb  X}}_L=\bar{\Bbb X}_Le^{\Bbb V_L},   \,\,\,\,  \Hat{\Bbb X}_L = \Bbb X_L, \,\,\,\, \Hat{ \Bbb  X}_R = e^{-\Bbb V_L} e^{i \bar {\tilde{ \Bbb V}}} \Bbb  X_R, \,\,\,\, \bar{ \Hat{\Bbb  X}}_R = \bar{\Bbb  X}_Re^{- i \tilde{\Bbb V} }\,, \nonumber
\eeq
in terms of which the Lagrangian  (\ref{Gauge action for EH}) reads (relabeling the fields $\hat{\Bbb X}_{L,R} \rightarrow \Bbb X_{L,R}$)
\beq
\mathcal{L}&=&\int d^2 \theta Q_+ Q_- \Big[ \bar{ \Bbb  X}_L \Bbb  X_L + \bar{ \Bbb  X}_R \Bbb X_R +\alpha (\bar{ \Bbb  X}_L \Bbb  X_R + \bar{ \Bbb  X}_R \Bbb  X_L)\Big] \\ \nonumber
&+&\int d^2\theta \left[ t \Bbb F - s \tilde{\Bbb  F} - c.c. \right]\,,
\eeq
where $d^2 \theta$ is the $\mathcal{N}=(1,1)$ measure and the relative minus sign between $s$ and $t$ comes from the ordering in the measure. Next, one imposes the fields to be gauge-covariantly semichiral and defines components with gauge-covariant $Q_{\pm}$'s, $i.e.$, 
\beq
X_L = \Bbb X_L\big|, \,\,\,\, Q_+ \Bbb X_L = i \mathcal{D}_+ \Bbb X_L, \,\,\,\, Q_- \Bbb X_L \big| = \Psi_{-}\,, \\  
X_R =\Bbb X_R \big|, \,\,\,\, Q_- \Bbb X_R = i \mathcal{D}_-  \Bbb X_R, \,\,\,\, Q_+ \Bbb X_R \big| = \Psi_{+}\,.
\eeq
The reduction of the semichiral vector multiplet  is given by \cite{Lindstrom:2007vc, Lindstrom:2008hx} 
\be
d^1=\left .\left( \Bbb F+\bar{\Bbb F} \right) \right|\,, \quad  d^2=\left. \left( \tilde{\Bbb F}+\bar{\tilde{\Bbb F}} \right)\right| \,, \quad   d^3=i\left. \left(  \Bbb F -\bar{\Bbb F} -\tilde{\Bbb F}+\bar{\tilde{\Bbb F}} \right)\right|\,,   \quad  f=-i \left. \left(  \Bbb F -\bar{\Bbb F} +\tilde{\Bbb F}-\bar{\tilde{\Bbb F}} \right)\right|\,.
\ee
Rescaling $X_L \rightarrow \alpha/(\sqrt{4-\alpha^2}) X_L$ and writing
\beq \nonumber
X_L = \frac{1}{4}\left(\frac{\phi_-}{\sqrt{\alpha+1} } - \frac{\bar \phi_+}{\sqrt{\alpha-1}}\right)\,, &&
X_R = \frac{1}{4}\left (\frac{\phi_-}{\sqrt{\alpha+1} } + \frac{\bar \phi_+}{\sqrt{\alpha-1}}\right)\,,
\eeq
the kinetic terms are diagonalized, $i.e.$, 
\beq \label{free action}
\mathcal{L}_{kin.}&\sim& \int d^2 \theta \left[    \mathcal{D}_+ \bar \phi_+ \mathcal{D}_- \phi_+ +  \mathcal{D}_+ \bar \phi_- \mathcal{D}_- \phi_- \right]
\eeq
and the constraints read
\beq \label{Constraints EH}
\bar \phi_+ \phi_+ - \bar \phi_- \phi_- &=& p\, , \\ \nonumber
\phi_+ \phi _- +i b&=&0\,, \nonumber
\eeq
where we have defined  
\beq
r\equiv -2 \text{Re}[s+t]\,, \qquad q \equiv -2 \text{Im}[t]\,, \qquad p \equiv -2 \text{Im}[s]\,,
\eeq
and $ 2b\equiv (r  +i q) \sqrt{\alpha^2-1} $. The free action (\ref{free action}), subject to the constraints (\ref{Constraints EH}), is the usual hyperk\"{a}hler quotient construction for Eguchi-Hanson \cite{Lindstrom:1983rt} (see also, $e.g.$, \cite{Curtright:1979yz,Rocek:1980kc}). This is a specific example of our discussion in Section \ref{Geometrical interpretation} of the semichiral quotient reducing to the hyperk\"{a}hler quotient. Thus, performing the quotient at the $\mathcal{N}=(2,2)$ level will give the generalized potential for this manifold. 

\subsection{Generalized Potential}

We have learned that the semichiral quotient (\ref{Gauge action for EH}) coincides, in  $\mathcal{N}=(1,1)$ language, to the hyperk\"{a}hler construction of Eguchi-Hanson. Therefore, performing the quotient in terms of $\mathcal{N}=(2,2)$ superfields will lead us to the generalized description of this manifold.  From (\ref{Gauge action for EH}), the equations of motion for the vector multiplet read
\beq \nonumber 
 e^{V_L}\left(1+|\Bbb  X_L|^2\right) 
+\frac{\alpha}{2}\left[e^{-i \tilde V} \left( 1+\bar{\Bbb  X}_R \Bbb  X_L\right)+  e^{i \bar {\tilde V} }\left(1+\bar{\Bbb  X}_L \Bbb  X_R  \right)\right]-\frac{(p+q)}{2}=0\,,\\ 
e^{V_R}\left(1+|\Bbb  X_R|^2\right)
 +\frac{\alpha}{2}\left[ e^{-i \tilde V} \left( 1+ \bar{\Bbb  X}_R \Bbb  X_L\right)+ e^{i \bar {\tilde V} }\left(1+ \bar{\Bbb  X}_L \Bbb  X_R\right) \right] -\frac{(p-q)}{2}=0\,,\\  \nonumber
\frac{i \alpha}{2} \left[ e^{-i \tilde V} \left(\bar{\Bbb  X}_R \Bbb  X_L+1\right)- e^{i \bar {\tilde V} }\left( \bar{\Bbb  X}_L \Bbb  X_R+1\right) \right] -\frac{r}{2}=0\,,
\eeq
where we have chosen the gauge $\Bbb X_L^{(2)} =\Bbb X_R^{(2)}=1$ (and relabeled the remaining fields). These  can be easily solved for $V_L,V_R, V'$, leading to the quotient potential
\beq \label{KEH} \nonumber
\hat K_{EH}&=&-\frac{p}{2} \log\left(\frac{-\left(q^2+r^2\right) (S^2- \alpha^2 T^2 )+p^2 \left(S^2+T^2 \alpha ^2\right)
-2 i p Q}{\left(S^2- \alpha ^2 T^2\right)^2}\right)  \\ 
&&-\frac{q}{2} \log\left(\frac{(1+|\Bbb  X_R|^2)^2 \left(p^2 S^2+r^2 S^2-q^2 \left(S^2-2 T^2 \alpha ^2\right)+2 i q Q\right)}{\left((p-q)^2+r^2\right) S^4}\right)  \\ \nonumber
&&- \frac{i r}{2} \log\left(\frac{(1+\bar{ \Bbb  X}_L \Bbb  X_R)^2 \left(-2 r^2 S^2+\left(p^2-q^2+r^2\right) T^2 \alpha ^2-2 r Q\right)}{T^4 \alpha ^2}\right)\,,
\eeq
where we have defined
 \beq
S^2 = (1+|\Bbb  X_L|^2 )(1+|\Bbb  X_R|^2 )\,, &&  T^2 = (1+\bar{ \Bbb  X}_R \Bbb  X_L )(1+ \bar{ \Bbb  X}_L \Bbb  X_R  )\,,
\eeq
and
\beq 
Q^2 =r^2 S^4 - (p^2 - q^2 + r^2) S^2 T^2 \alpha^2 - q^2 T^4 \alpha^4\,.
\eeq
This quotient construction has been discussed to some extent in \cite{Merrell:2007sr}, where the authors suggest that this will lead to a non-trivial H-flux. Based on our result of Section \ref{The semichiral Quotient}, we know this is not the case. Instead, it must describe a hyperk\"{a}hler manifold; in this case, Eguchi-Hanson. By explicit calculation, one can also verify that  (\ref{KEH}) satisfies the Monge-Ampere equation (\ref{c}) with $c=2(1-\frac{2}{\alpha^2})$, $i.e.$, 
\beq
\{J_+,J_-\}= 2(1-\frac{2}{\alpha ^2}) \Bbb I\,.
\eeq
To show explicitly that one can derive the standard metric for Eguchi-Hanson from this potential, we set the FI parameters to some convenient value for which the potential is simplified. The choice $r=q=0$, for instance, leads to the left-right symmetric potential
\beq \label{KEHp}
K=p \log\left[S+\alpha T \right]\,,
\eeq
while the choice $r=0,\,p=-q$ leads to
\beq
 K=p \log \left[ \frac{S^2-\alpha^2 T^2}{1+|\Bbb  X_L|^2}\right]\,. 
 \label{EH Potential}
\eeq
This form of the potential also coincides with that of \cite{arXiv:1111.3893}, constructed by twistor methods.  Note how these potentials are more compact than the usual K\"{a}hler potential and contain no term with a square root outside the $\log$.  Working with the potential (\ref{EH Potential}), we will explicitly construct the Eguchi-Hanson metric, but first we will study the $SU(2)$ symmetry of the problem. 

\subsection{$SU(2)$ symmetry}

The action (\ref{Gauge action for EH})  is invariant under the global $SU(2)$ transformations which rotate $(\Bbb X^{(1)}, \Bbb X^{(2)})$, as well as under $U(1)$ gauge transformations. Recall that we have chosen the $U(1)$ gauge 
\be \label{gauge semis}
\Bbb X_L^{(2)}=\Bbb  X_R^{(2)}=1\,,
\ee
which is not preserved by the $SU(2)$. Nevertheless,  the $SU(2)$ symmetry  can be realized non-linearly in the gauged action by introducing a compensating $U(1)$ transformation with parameter $\Lambda_C$, namely

\beq
\left( \begin{array}{cc} \delta \Bbb X_L^{(1)} \\ \delta \Bbb  X_L^{(2)}  \end{array} \right)= i \left( \begin{array}{cc} \alpha & -i \beta \\ i \bar \beta & -\alpha  \end{array} \right) \left( \begin{array}{cc}  \Bbb  X_L^{(1)} \\ \Bbb  X_L^{(2)}  \end{array} \right) +i  \left( \begin{array}{cc} \Lambda_C \Bbb X_L^{(1)} \\ \Lambda_C \Bbb X_L^{(2)}  \end{array} \right)\,,
\eeq 
and similarly for $\Bbb  X_R$. Imposing that the transformation preserves the gauge (\ref{gauge semis}), and relabelling $\Bbb{ X}_{L,R}^{(1)}=\Bbb {X}_{L,R}$ henceforth, one finds
\beq
\delta \Bbb  X_L=2 i \alpha \Bbb  X_L+\bar \beta  (\Bbb X_L)^2+\beta\,, &&
\delta \Bbb  X_R=2 i \alpha \Bbb X_R+\bar \beta  (\Bbb X_R)^2+\beta\,.
\eeq
The infinitesimal transformations are generated by the vector field
\beq
\xi= \delta \Bbb X_L \partial_{L} +\delta \Bbb X_R \partial_{R} + c.c.
\eeq
and the finite transformations are given by the M\"{o}bius transformations
\beq
\Bbb X_L\rightarrow \frac{a \Bbb X_L +b}{\bar a - \bar b \Bbb X_L}\,, && \Bbb X_R\rightarrow \frac{a \Bbb X_R +b}{\bar a - \bar b \Bbb X_R}\,,
\eeq
with $|a|^2+|b|^2=1$. Given the $SU(2)$ invariance of the potential (and therefore the metric), it will be convenient to find coordinates in which this symmetry is manifest. The first step in doing this is to note that a natural radial coordinate $R$ is defined by the invariant cross-ratio
\beq
R^2\equiv \frac{Z_{13}Z_{24}}{Z_{23}Z_{14}}\,,
\eeq
where $Z_{ij}=Z_i-Z_j$. Since we have only two complex variables, namely $X_L, X_R$, there is only one, non-zero, independent cross ratio we can form. Taking $Z_1=X_L, Z_2= X_R, Z_3=-1/\bar{X}_L$ and  $Z_4=-1/ \bar{X}_R$ we have
\beq
R^2 = \frac{(1+| X_L|^2)(1+| X_R|^2)}{(1+ \bar{X}_L X_R)( 1+ \bar{ X}_R  X_L)} = \frac{S^2}{T^2}\,.
 \label{cross-ratio}
\eeq
One can easily verify that $\mathcal{L}_{\xi} R = \xi  R =  0$.  Therefore, one can reach every point $(X_L,X_R)$, at a certain radius $R$, by  choosing a point $(X_L^0,X_R^0)$ on the sphere of that radius and acting by a $SU(2)$ transformation with parameters $(a,b)$. Thus, we can parameterize any point $(X_L,X_R)$ by $a,b$ (subject to $|a|^2+|b|^2=1$), and the radial coordinate $R$. Then, the natural remaining invariants are the Cartan 1-forms $\sigma^i$ on the group manifold.  As shown in Appendix \ref{AppendixSU2}, this parameterization of the $X_L,X_R$ coordinates leads to 
\beq \label{dx Cartan forms}\nonumber
dX_L &=& \frac{1}{ \bar a ^2}(i \sigma^1 - \sigma^2)\,, \\ 
dX_R&=&\frac{1}{(\bar a - \rho \bar b)^2}\left[2 i \rho \sigma^3 + i (1-\rho^2) \sigma^1 - (1+\rho^2) \sigma^2+ d \rho \right]\,,
\eeq
where  $\rho^2 \equiv R^2-1$. As we shall see below, when writing the line element in this $SU(2)$ parameterization, all the dependence in $a,b$ drops out as a consequence of the invariance of the metric.
Also, one can see by explicit calculations of  $J_{\pm}$ from the potential (\ref{KEHp}) that
\beq
\mathcal{L}_{\xi}J_{\pm}=0\,.
\eeq
That is, both complex structures, $J_{\pm}$ (and therefore the third one), are preserved by the $SU(2)$, which is an important property of Eguchi-Hanson (see Appendix \ref{AppendixSU2} for more details). To show explicitly that the potential (\ref{EH Potential}) indeed describes this manifold, we compute the metric.

\subsection{Metric}

From the potential (\ref{EH Potential}), and Eqs. (\ref{G and B})-(\ref{K}), one finds the metric\footnote{For simplicity, we have taken $\alpha=\sqrt 2$ here, but the final result (\ref{EH metric}) holds for any $\alpha$, with appropriate redefinitions.} 

\beq \nonumber
ds^2&=& \frac{F(R) (\bar X_L-\bar X_R)^2}{(1+\bar X_R X_L)^2(1+|X_L|^2)^2} d X_L d X_L+ \frac{F(R) (\bar X_R-\bar X_L)^2}{(1+\bar X_L X_R)^2(1+|X_R|^2)^2} d X_R d X_R  \\ 
&& + \frac{G(R)}{(1+|X_L|^2)^2} d \bar X_L d X_L + \frac{G(R)}{(1+|X_R|^2)^2} d \bar X_R d X_R \\ \nonumber
&& + \frac{H(R) (\bar X_L- \bar X_R)^2 }{(1+|X_L|)^2(1+|X_R|^2)^2} d X_L  d X_R + \frac{I(R)(1+\bar X_L X_R)^2}{(1+|X_L|)^2(1+|X_R|^2)^2} d X_L d \bar  X_R \\  \nonumber
&&+c.c.
\eeq
where
\beq \nonumber
F(R)= -\frac{16 \left(2-2 R^2+R^4\right) }{\left(-2+R^2\right)^3 R^2}\,,  && G(R)=-\frac{8 (2 - 2 R^2 + R^4)^2}{(-2 + R^2)^3 R^2}\,,  \\ 
H(R)=\frac{4 R^2 \left(4-2 R^2+R^4\right) }{\left(-2+R^2\right)^3}\,, && I(R)= \frac{4 R^2 \left(4-6 R^2+3 R^4\right)}{\left(-2+R^2\right)^3}\,.  \\ \nonumber
\eeq
Defining $r$ through 
\beq
R^2=\frac{2 r^2}{r^2-1}\,,
\eeq
and using (\ref{dx Cartan forms}), after some algebra the line element reads
\beq \label{EH metric}
\frac{1}{8} ds^2=\frac{1}{1-\frac{1}{r^4}}dr^2+r^2 \left( \sigma_1^2 +\sigma_2^2 + (1-\frac{1}{r^4}) \sigma_3^2 \right)\,,
\eeq
which is the usual Eguchi-Hanson metric (see, $e.g.$, \cite{Eguchi:1980jx} for a review).

\section{Taub-NUT}
\label{Taub-NUT}

\subsection{A gauged linear sigma model}

Here we present a gauged linear sigma model in terms of semichiral superfields whose low-energy limit target space is Taub-NUT. Consider a gauged linear sigma model with two copies of semichiral superfields, just as the Eguchi-Hanson case, but with the difference that the isometry acts by translations on one of the pairs, $i.e.$,

\beq \nonumber
 \label{gauge theory Taub-NUT}
K&=& \Bbb{\bar{X}}_L^{(1)} e^{V_L} \Bbb  X_L^{(1)} +\Bbb{\bar{X}}_R^{(1)} e^{V_R}  \Bbb X_R^{(1)} +\alpha ( \Bbb{\bar{X}}_R^{(1)} e^{-i \tilde{\Bbb  V}} \Bbb  X_L^{(1)}+\Bbb{ \bar{X}}_L^{(1)} e^{i \bar{\tilde{ \Bbb V}}} \Bbb  X_R^{(1)}) \\ 
&&+\frac{1}{2} \left ( \Bbb{X}_L^{(2)}+ \Bbb{\bar X}_L^{(2)}+ V_L \right)^2+\frac{1}{2}  \left ( \Bbb{X}_R^{(2)}+ \Bbb{\bar X}_R^{(2)}+ V_R\right)^2 \\  \nonumber
&&+ \frac{\alpha}{2} \left(  ( \Bbb{X}_L^{(2)}+ \Bbb{\bar X}_R^{(2)}- i \tilde{\Bbb{V}} )^2+ ( \Bbb{X}_R^{(2)}+ \Bbb{\bar X}_L^{(2)}+ i \bar{\tilde{\Bbb{V}}})^2 \right)  \\  \nonumber
&&- (t \Bbb V +s \tilde{ \Bbb V} +c.c.)\,.
\eeq
It is known in general that such constructions (where the isometry acts transitively on some fields) lead to ALF (as opposed to ALE) spaces and we claim that performing the semichiral quotient in this way leads to the semichiral description of Taub-NUT. Although integrating out the vector field cannot be done explicitly, by implicit differentiation we could still compute  the metric. Instead, we shall study the geometry of the T-dual  theory. 

\subsection{T-dual}

To perform a T-duality from the worldsheet perspective, one proceeds as according to \cite{Lindstrom:2007sq, Rocek:1991ps}. We introduce an additional vector multiplet $U^{\alpha}$, which acts on the second pair and constrain its field strengths to be trivial by Lagrange multipliers $\Phi,\chi$, $i.e.$, 

\beq \nonumber
\tilde K&=& \Bbb{\bar{X}}_L^{(1)} e^{V_L} \Bbb  X_L^{(1)} +\Bbb{\bar{X}}_R^{(1)} e^{V_R}  \Bbb X_R^{(1)} +\alpha \left ( \Bbb{\bar{X}}_R^{(1)} e^{-i \tilde{\Bbb  V}} \Bbb  X_L^{(1)}+\Bbb{ \bar{X}}_L^{(1)} e^{i \bar{\tilde{ \Bbb V}}} \Bbb  X_R^{(1)} \right) \\ 
&&+\frac{1}{2} \left ( \Bbb{X}_L^{(2)}+ \Bbb{\bar X}_L^{(2)}+ U_L \right)^2+\frac{1}{2}  \left ( \Bbb{X}_R^{(2)}+ \Bbb{\bar X}_R^{(2)}+ U_R\right)^2 \\  \nonumber
&&+ \frac{\alpha}{2} \left(  ( \Bbb{X}_L^{(2)}+ \Bbb{\bar X}_R^{(2)}- i \tilde{\Bbb{U}} )^2+ ( \Bbb{X}_R^{(2)}+ \Bbb{\bar X}_L^{(2)}+ i \bar{\tilde{\Bbb{U}}})^2 \right)  \\  \nonumber
&&- ((t+\Phi) \Bbb V +(s+\chi) \tilde{ \Bbb V} +c.c.) +(\Phi \Bbb U +\chi \tilde{\Bbb U} +c.c.)\,,
\eeq
were we have shifted $U^{\alpha}\rightarrow U^{\alpha}-  V^{\alpha}$. Integrating out $ U^{\alpha}$ yields the T-dual gauged linear sigma model
\beq \label{K for NS5} \nonumber
\tilde K&=& \frac{1}{g^2}(- \frac{\bar \chi \chi}{\alpha^2-1} + \bar \Phi \Phi)   + \bar{\Bbb X}_L e^{V_L}\Bbb  X_L +\bar{\Bbb X}_R e^{V_R}  \Bbb X_R +\alpha (\bar{\Bbb X}_R e^{-i \tilde{\Bbb  V}}\Bbb X_L+\bar{\Bbb X}_L e^{i \bar{\tilde{ \Bbb V}}} \Bbb X_R   ) \\ 
 &&-(\Phi \Bbb V +\chi \tilde{ \Bbb V} +c.c)\,, 
\eeq
where we have shifted $\chi, \phi$ to get rid of the FI parameters $s$ and $t$, dropped terms that vanish upon integration in superspace ($i.e.$, generalized K\"{a}hler transformations) and rescaled the fields appropriately\footnote{We have chosen to keep the kinetic terms of $\Phi$ with the usual normalization, leading to the $1/(\alpha^2-1)$ factor for $\chi$. This relative coefficient, as we will see, is important to ensure the $\mathcal{N}=(4,4)$ symmetry of the quotient model, as is expected of a model which is dual in this manner to a model describing a hyperk\"{a}hler manifold in terms of semichiral fields, as discussed in Section \ref{T-Duality}. }.  As we will see in Section \ref{NS5 Branes}, this gauged linear sigma model describes a smeared NS5-brane and, therefore, the original theory (\ref{gauge theory Taub-NUT}) is a gauged sigma model  for Taub-NUT. 

\section{NS5-branes}
\label{NS5 Branes}

It is well known that under type II string theory T-duality, Taub-NUT is mapped to an NS5-brane. A worldsheet discussion of such relation is given in  \cite{Tong:2002rq}. There, a gauge theory description of NS5-branes involving a hypermultiplet, a twisted hypermultiplet, and a vector multiplet acting on the former is given and instanton corrections are discussed. We shall first show that the gauge theory  (\ref{K for NS5}), involving semichiral fields, also describes NS5-branes and we shall comment in Section \ref{Worldsheet Instantons} on instanton effects.  \\
\subsection{A gauged linear sigma model}
Consider the action

\beq  \nonumber  \label{actionNS5}
\mathcal{L}&=&\int d^4 \theta \Big[  \frac{1}{e^2} ( \Bbb F  \bar{\Bbb F}  -\Bbb{ \tilde F}  \Bbb{ \bar{\tilde F}}    )+ \frac{1}{g^2}(- \frac{\bar \chi \chi}{\alpha^2-1} + \bar \Phi \Phi)   \\ 
&& \, + \bar{\Bbb  X}_L e^{V_L} \Bbb  X_L +\bar{\Bbb X}_R e^{V_R}  \Bbb X_R +\alpha (\bar{\Bbb X}_R e^{-i \tilde{\Bbb  V}}\Bbb X_L+\bar{\Bbb  X}_L e^{i \bar{\tilde{ \Bbb V}}} \Bbb  X_R   ) \\   \nonumber
&&-( \Phi \Bbb V  +\chi \tilde{ \Bbb V} +c.c.)  \Big]\,.
\eeq
In the IR limit ($ e^2 \rightarrow \infty$), the equations of motion for the semichiral vector field are

\beq \nonumber
\bar{\Bbb  X}_L e^{V_L}\Bbb X_L +\frac{\alpha}{2}( \bar{\Bbb X}_R e^{-i \tilde V}\Bbb X_L+ \bar{\Bbb X}_L e^{i \bar {\tilde V} }\Bbb X_R ) -\frac{i}{2} (\Phi-\bar \Phi + \chi-\bar \chi)=0\,,\\
\bar{\Bbb X}_R e^{V_R}\Bbb X_R +\frac{\alpha}{2}( \bar{\Bbb X}_R e^{-i \tilde V}\Bbb X_L+ \bar{\Bbb X}_L e^{i \bar {\tilde V} }\Bbb X_R ) -\frac{i}{2} (-\Phi+\bar \Phi+\chi-\bar \chi)=0\,,\\  \nonumber
\alpha \frac{i}{2} ( \bar{\Bbb X}_R e^{-i \tilde V}\Bbb X_L- \bar{\Bbb X}_L e^{i \bar {\tilde V} }\Bbb X_R) +\frac{1}{2}(\chi+\bar \chi +\Phi+\bar \Phi)=0\,. 
\eeq
For simplicity, we gauge the semis to $\Bbb X_L=\Bbb X_R=1$. Solving these equations leads to 
\beq 
K(\Phi, \chi)&=& \frac{1}{g^2}(- \frac{\bar \chi \chi}{\alpha^2-1} + \bar \Phi \Phi)   +\Delta K(\Phi, \chi) 
\eeq
with
\beq \nonumber
\Delta K(\Phi, \chi)&\equiv&\, - i \chi \log\Big[ i (\chi -\bar \chi)\alpha^2 +i (\chi +\bar \chi +\Phi +\bar \Phi)(\alpha^2-1)-2 R \Big]\\ 
&&\, - i \Phi \log\Big[ - \frac{i (\chi +\bar \chi +\Phi +\bar \Phi) + i (\Phi -\bar \Phi) \alpha^2 +2 R}{2 i (\Phi +\bar \chi)}\Big]+c.c.\,, \label{Delta K}
\eeq
where we have defined
\beq
R\equiv \frac{1}{2} \sqrt{(\chi +\bar \chi +\Phi +\bar \Phi)^2(\alpha^2-1)-(\chi -\bar \chi) ^2 \alpha^2 -(\Phi -\bar \Phi)^2 \alpha^2(\alpha^2-1)}\,.
\eeq
Note that $\alpha^2 \geq 1$ ensures the reality of $R$. From here we find
\beq \nonumber
K_{\chi \bar \chi}&=&-\frac{1}{\alpha^2-1}\Big(\frac{1}{g^2}+\frac{\alpha^2-1}{2 R} \Big)\,, \,\,\,\,\, K_{\Phi \bar \Phi} =\frac{1}{g^2}+\frac{\alpha^2-1}{2 R}\,,  \\ 
K_{\chi \bar \Phi}&=&-\frac{1}{2 R} \Big(\frac{(\alpha^2-1)(\bar \Phi +\chi)}{2 i R -(\chi -\bar \chi) -(\alpha^2-1)(\Phi-\bar \Phi)}\Big)\,.
\eeq
Note that the $1/(\alpha^2-1)$ factor for $\bar \chi \chi$ in (\ref{actionNS5}) is crucial for the potential to satisfy the scaled Laplace equation (\ref{Scaled Laplace Equation}) (although in Section \ref{T-Duality} we performed the duality in the other direction, one would expect the same relations to hold). After a trivial rescaling of the fields, the line element is given by
\beq
ds^2=2(K_{\Phi \bar \Phi} d \Phi d\bar \Phi -K_{\chi \bar \chi} d\chi d \bar \chi)=2 H(r)( d \Phi d\bar \Phi + d\chi d \bar \chi) 
\label{metric NS5}
\eeq
with
\beq
H(r)\equiv \Big(\frac{1}{g^2}+\frac{1}{2 r } \Big)\,. 
\eeq
Defining
\beq
\chi | = \frac{ \left( r_1 +\theta \right)}{2}+ i \frac{r_2}{\sqrt 2}\,, && \Phi| = \frac{ \left( r_1 -\theta \right)}{2 } + i \frac{r_3}{\sqrt 2}\,, \\  \nonumber
\eeq
we finally have
\beq
ds^2=  H(r)(  d \vec  r \cdot  d \vec  r+ d\theta^2)\,,
\eeq
which is the metric for an NS5-brane, smeared along the $\theta$ direction. 

\subsection{Comment on instanton corrections}
\label{Worldsheet Instantons}

In \cite{Tong:2002rq} a gauge theory description of smeared NS5-branes and a worldsheet T-dual description of Taub-NUT was also given.  It was argued that worldsheet instanton corrections to the effective action  un-smear the NS5-brane, localizing it in the $\theta$ direction. (For a recent discussion of this phenomenon in the context of double field theory \cite{Hull:2009mi}, see \cite{Jensen:2011jn}.) In two dimensions, instantons are Nielsen-Olesen vortices, which arise as BPS solutions to an abelian Higgs model contained in the gauge theory. Although our  gauge theory construction is quite different (from the $\mathcal{N}=(2,2)$ point of view), the same arguments hold so we expect the same mechanism to be at work.  Our construction does not add to the results of  \cite{Tong:2002rq}, but is consistent with it. This is more easily seen upon reduction of the gauge theory  (\ref{actionNS5}) to $\mathcal{N}=(1,1)$. Following \cite{Lindstrom:2007vc}, we get (see Appendix \ref{AppendixN=1} for details)

\beq \nonumber
\mathcal{L}&=& \int d^2 \theta \Big [ \frac{1}{4 e^2}( D_+ d^{a}) \, (D_- d^{b})\, g_{ab} +  \frac{1}{g^2}\left(  D_+ \bar \phi \, D_- \phi + D_+ \bar \chi \, D_- \chi     \right) +(\mathcal{D}_+ X^i)(\mathcal{D}_- X^j)E_{ij}   \\ \nonumber
&&+2i d^1( \bar X_L X_L - \bar X_R X_R -\frac{i}{8}(\phi -\bar \phi)) 
+ d^3( \alpha(\bar X_R X_L -\bar X_L X_R) -\frac{i}{8} (\phi +\bar \phi +\chi +\bar \chi)) \\ \nonumber
&&- 2i d^2( \bar X_L X_L +\bar X_R X_R +\alpha (\bar X_R X_L +\bar X_L X_R) -\frac{i}{8} (\chi -\bar \chi)) + i f(\phi +\bar \phi -\chi -\bar \chi)\Big ]\,,
\eeq
where $d^a=(f,d^1, d^2,d^3)$, $X^i=(X_L, \bar{X}_L,X_R,\bar X_R)$ and $g_{ab}=diag(1,2,2,1)$. One can rewrite this in terms of the fields $\phi_{\pm}$ from Section \ref{reduction to N=1} which diagonalize the kinetic terms for the semis. Then, following Tong, we allow only the lowest component of, say, $\phi_+$ to vary over space and set all other fields to their classical expectation values. This results in an abelian Higgs model with a $\theta$ term for the gauge field, whose instanton solutions (in the limit $g^2\rightarrow0$) are conjectured to contribute to the low-energy effective action, effectively replacing 
\beq
H(r)\rightarrow H(r,\theta)= \frac{1}{g^2} +\frac{1}{2r} \frac{\sinh r}{\cosh r -\cos{\theta}}\,,
\eeq 
therefore unsmearing the NS5.  

\section{T-dual of Eguchi-Hanson}
\label{T-dual of Eguchi-Hanson}

For completeness, we finally discuss the T-dual of Eguchi-Hanson. We can perform a T-duality before taking the quotient. As before, we introduce an additional semichiral vector multiplet $U^{\alpha}$ which acts only on the second pair $\Bbb X^{(2)}_{L,R}$,  and defines
\beq \nonumber
K&=& \left( \bar{\Bbb  X}_L^{(1)} \Bbb  X_L^{(1)}+ \bar{ \Bbb  X}_L^{(2)} \Bbb X_L^{(2) }e^{U_L}  \right)e^{V_L}+ \left( \bar{ \Bbb X}_R^{(1)} \Bbb X_R^{(1)}+ \bar{ \Bbb X}_R^{(2)} \Bbb X_R^{(2) } e^{U_R}  \right)e^{V_R} \\ 
&& + \alpha \left( \Bbb  X_L^{(1)} \bar{ \Bbb X}_R^{(1)} +   \Bbb X_L^{(2)} \bar{\Bbb  X}_R^{(2)} e^{-i \tilde{U}}\right)e^{-i \tilde{V}}+\alpha \left( \bar{\Bbb  X}_L^{(1)}\Bbb X_R^{(1)} +   \bar{ \Bbb X}_L^{(2)}  \Bbb X_R^{(2)} e^{i \bar{\tilde{U}}}\right)e^{i \bar{\tilde{ V}}} \\ \nonumber
&&-\left[\Phi \Bbb U +\chi \tilde{ \Bbb  U} +c.c.\right]\,.
\eeq
Shifting $U^{\alpha}\rightarrow U^{\alpha}-V^{\alpha}$, the Lagrangian decouples and, gauging all the semis to 1, we have
\beq
K&=&K_{1}+K_{2} \,,
\eeq
where
\beq \nonumber
K_{1}&=&e^{U_L}+e^{U_R} +\alpha (e^{-i \tilde{\Bbb  U}}+ e^{i \bar{\tilde{ \Bbb U}}} ) +\left(\Phi \, \Bbb U +\chi \, \tilde{ \Bbb U}+c.c\right), \\ \nonumber
K_{2}&=&e^{V_L}+e^{V_R} +\alpha (e^{-i \tilde{\Bbb  V}}+ e^{i \bar{\tilde{ \Bbb V}}} ) -\left((\Phi+t) \Bbb V +(\chi+s) \tilde{\Bbb V}+c.c\right)\,. 
\eeq
Thus, integrating out both $U^{\alpha}$ and $V^{\alpha}$ reduces to the case studied for NS5-branes with $K_{1} =\Delta K(-\Phi,-\chi)$\,,
$K_{2} =\Delta K(\Phi+t,\chi+s)$ and therefore 
\beq
\tilde{K}= \Delta K(-\Phi,-\chi)+\Delta K(\Phi+t,\chi+s)\,,
\eeq
with $\Delta K$ given in (\ref{Delta K}). Since the metric is linear in second derivatives of the potentials, we have
\beq
\tilde{K}_{\chi \bar \chi}=-\frac{1}{2}\Big(\frac{1}{ R_1}+\frac{1}{ R_2} \Big)\,, &&  \tilde{K}_{\Phi \bar \Phi} =\frac{\alpha^2-1}{2}\Big(\frac{1}{ R_1}+\frac{1}{ R_2} \Big)\,,
\eeq
and similarly for the torsion terms. Again, this potential satisfies the scaled Laplace equation
\beq
\tilde{K}_{\Phi \bar \Phi}+(\alpha^2-1)\tilde{K}_{\chi \bar \chi}=0\,, 
\eeq
in accordance with our results of Section \ref{T-Duality}. Note that changing the relative position of the mass-points corresponds to rotating the complex structures.

\section{Summary and Conclusions}

We have studied a supersymmetric quotient construction by the use of general $\mathcal{N}=(2,2)$ sigma models and the semichiral vector multiplet. We first restricted ourselves to the case in which only semichiral fields are involved. Due to the presence of a $b$-field in these models, one may naively think that a non-zero H-flux could be induced on the quotient manifold $\mathcal{\hat M}$, even if the original manifold $\mathcal{M}$ is hyperk\"{a}hler.  This, however, is prevented by our result of Section \ref{The semichiral Quotient}, asserting that the quotient of a hyperk\"{a}hler manifold is hyperk\"{a}hler, as in the usual hyperk\"{a}hler quotient. Furthermore, the value of the anticommutator of the complex structures is preserved under the studied quotient. Thus, although the quotient manifold in general does have a $b$-field, its field strength $H=db$ vanishes. Nonetheless, the quotient provides a powerful method for constructing generalized potentials for hyperk\"{a}hler manifolds, of which few explicit examples are known. We gave two examples of well-known hyperk\"{a}hler manifolds, namely Eguchi-Hanson and Taub-NUT. We also used the SVM to perform T-duality transformations, giving a new $\mathcal{N}=(2,2)$  gauged linear sigma model description of  (smeared) NS5-branes involving semichiral, chiral, and twisted-chiral superfields. This description is consistent with previous ones in that it contains an abelian Higgs model whose instanton solutions unsmear the NS5.

We have also clarified and extended some previous results on the duality relation of these semichiral models with $\mathcal{N}=(4,4)$ models for chiral/twisted-chiral fields. We showed that the T-dual of an $\mathcal{N}=(4,4)$ model for chiral/twisted-chiral fields, may or may not describe a hyperk\"{a}hler manifold, depending on the character of the isometry along which the duality is performed. If the isometry is translational, the dual manifold is hyperkh\"{a}ler. For a general isometry, however, the dual manifold is in general not hyperk\"{a}hler, even if the $\mathcal{N}=(4,4)$  SUSY is preserved. This, for instance, is the case of the $SU(2)\times U(1)$ WZW model which was briefly discussed.

 We also commented on more general quotients that can lead to manifolds with torsion, noting that this requires the presence of more than one type of $\mathcal{N}=(2,2)$ field and gave an example involving a chiral and a pair of  semichiral fields. A more thorough analysis of such general quotients remains open.\\

\begin{large}{\bf Acknowledgements:}\end{large}\\
I especially wish to thank Martin Ro\v{c}ek for suggesting this project and for crucial guidance at all stages of its development. I also wish to thank  Malte Dyckmanns for valuable discussions and reviewing the manuscript, Malin G\"{o}teman for pointing out an important typo, Ulf Lindstr\"{o}m for reviewing the manuscript, and Alexander Sevrin and Warren Siegel for discussions. The research of PMC was supported by NSF grant no. PHY-0969739.

\appendix

\section{Semichiral Quotient}
\label{semichiral Quotient}
Here we give the necessary elements and sketch the proof of (\ref{c Hat}). As mentioned in the text, the requirement $\{J_+,J_-\}=c\, \Bbb I$ implies the set of equations
\beq   \label{c1 Appendix}
\left\{\mathcal{K}_{LR}^{-1}C_{RR}\mathcal{K}_{RL}^{-1},J_s\right\}&=&0\,,   \\[2mm]  \label{c2 Appendix}
J_s \mathcal{K}_{LR}^{-1}J_s \mathcal{K}_{RL}  +   \mathcal{K}_{LR}^{-1} J_s \mathcal{K}_{RL}J_s +\mathcal{K}_{LR}^{-1} C_{RR} \mathcal{K}_{RL}^{-1} C_{LL}&=& c\,  \Bbb{I}\,. 
\eeq
We define the potential $\hat K$ by
\beq
\hat K(\Bbb X^{i}_{l},\Bbb X^{i'}_r)= K(\Bbb X^{i}_{l},\Bbb X^{i'}_r; \Bbb X^{\alpha}+V^{\alpha}) - t_{\alpha} V^{\alpha}\,,
\eeq
from where the standard relation of second derivatives
\beq
\hat K_{\mu \nu} = K_{\mu \nu} - K_{\mu \alpha} K_{\beta \alpha}^{-1} K_{\beta \nu}
\eeq
follows, where $\mu=(i,i',\bar i, \bar i')$ labels the $4N$ coordinates. From now on we suppress obvious indices, writing $(L,R)=(l,r,\alpha)$. Capital letters refer to the manifold $\mathcal{M}$, while lower-case are coordinates on $\hat{\mathcal{M}}$ and $\alpha$ labels coordinates which are gauged away. We decompose the relevant matrices as

\beq \nonumber
\mathcal{K}_{LR}=\left(
\begin{array}{c|c}
K_{lr} &  K_{l\alpha}\\ \hline
K_{\beta r} &  K_{\beta \alpha}
\end{array}\right), \,\,\, \mathcal{K}_{LL}=\left(
\begin{array}{c|c}
K_{ll} &  K_{l\alpha}\\ \hline
K_{\beta l} &  K_{\beta \alpha}
\end{array}\right), \,\,\,
 \,\,\, C_{LL}=\left(
\begin{array}{c|c}
C_{ll} &  C_{l\alpha}\\ \hline
C_{\beta l} &  C_{\beta \alpha}
\end{array}\right), \,\,\,
J_s= \left(
\begin{array}{c|c}
\hat J &  0\\ \hline
0 &  j
\end{array}\right)\,,
\eeq
with $\hat J^2=-1$ and $j^2=-1$ and 
\beq \nonumber
C_{ll}=[\hat J,K_{ll}], &&C_{\beta \alpha}=[j,K_{\beta \alpha}]\,, \\
C_{l \alpha} =  \hat J K_{l \alpha}- K_{l \alpha} j\,, && C_{\beta l} = j K_{\beta l} - K_{\beta l }\hat J, \label{def c's}
\eeq
(and similarly for $C_{RR}$).  The inverse matrices are given by
\beq
\mathcal{K}_{RL}^{-1}\equiv (\mathcal{K}_{LR})^{-1}= \left(
\begin{array}{c|c}
\hat{K}_{lr}^{-1} &  -\hat{K}_{lr}^{-1}K_{l \alpha} K_{\beta \alpha}^{-1}\\ \hline
- K_{\beta \alpha}^{-1} K_{\beta r} \hat{K}_{lr}^{-1} &  T^{\alpha \beta}
\end{array}\right)
\eeq
and $\mathcal{K}_{LR}^{-1}=( \mathcal{K}_{RL}^{-1})^t$ and where 
\beq \label{K hat}
\hat{K}_{lr} &=& K_{lr}-K_{l \alpha}K_{\beta \alpha}^{-1}K_{\beta r}, \\  \label{T}
T^{\alpha \beta}&=&K_{\beta \alpha}^{-1}+ K_{\delta \alpha}^{-1}K_{\delta r} \hat{K}_{lr}^{-1}K_{l \gamma}K_{\beta \gamma}^{-1}\,.
\eeq
(Here we have changed the notation slightly to mean $\hat K_{lr}^{-1}= (\hat K_{lr})^{-1}$, $K_{\beta \alpha}^{-1}=(K^{-1})^{ \alpha \beta}$, $etc.$). Similarly, we also have
\beq
\hat{C}_{rr}= C_{rr} -\left[ \hat J, K_{r \beta }K_{\alpha \beta}^{-1}K_{\alpha r} \right], && \hat{C}_{ll}= C_{ll} -\left[ \hat J, K_{l \beta }K_{\alpha \beta}^{-1}K_{\alpha l} \right]\,. 
\eeq
By rather straightforward (albeit tedious) algebraic manipulations, one can show that (\ref{c1 Appendix}) and  (\ref{c2 Appendix}) lead to 
\beq
\left\{\hat{K}_{rl}^{-1}\hat{C}_{rr}\hat{K}_{lr}^{-1},\hat{J}\right\}&=&0\,, \label{off-diagonal quotient}\\
\hat{J} \hat{K}_{rl}^{-1}\hat J \hat{K}_{rl}+\hat K_{rl}^{-1} \hat{J} \hat{K}_{rl} \hat J +\hat K_{rl}^{-1} \hat C_{rr} \hat{K}_{lr}^{-1} \hat{C}_{ll}&=& c\,  \Bbb{I}\,,
\eeq 
which is equivalent to the statement that
\beq
\{\hat{J}_+,\hat{J}_-\}=c\, \Bbb I\,,
\eeq
as we wanted to prove.

\section{T-duality}
 \label{T-duality appendix}

Here we give some of the details leading to (\ref{c translational Killing vector}) and (\ref{c General Killing vector }). Writing the Legendre transform (\ref{T-dual Sevrin}) as
\beq
K(\Bbb X^i)=F(V^{\alpha})-\frac{1}{2} V^{\alpha}\delta_{\alpha i } \Bbb X^i \,,
\eeq
where we defined
\beq \nonumber
\Bbb X^i \equiv \left(i(\Bbb X_L-\bar{ \Bbb X}_L +\Bbb X_R -\bar{\Bbb X}_R),- i(\Bbb X_L-\bar{\Bbb X}_L -\Bbb X_R +\bar {\Bbb X}_R),-( \Bbb X_L+\bar{\Bbb  X}_L +\Bbb X_R +\bar{\Bbb X}_R) \right)\,,
\eeq
we find the standard relation of second derivatives
\beq 
K_{i j} = -\frac{1}{2}\delta_{i \alpha} (F^{-1})^{\alpha  \beta}\delta_{ \beta j }\,.
\eeq
Explicitly inverting the general $3\times3$ matrix $F_{\alpha \beta}$ and using these relations in the definition (\ref{c in 4 dimensions}), one finds
\beq \label{c in terms of V}
c=\frac{2 \left(F_{V^{\phi}V^{\phi}}+F_{V^{\chi}V^{\chi}}+2 F_{V'V'}\right)}{F_{V^{\phi}V^{\phi}}-F_{V^{\chi}V^{\chi}}}\,.
\eeq
The important point now is that the Laplace equation ($F_{\phi \bar \phi}+F_{\chi \bar \chi}=0$) translates into 
\beq \label{rotated eq}
\cos^2(\theta)F_{V^{\phi}V^{\phi}}+\sin^2(\theta)F_{V^{\chi}V^{\chi}}+F_{V'V'}=0\,,
\eeq
which is a direct consequence of how the gauging was performed in (\ref{LVM gauging}) (\textit{i.e.}, the charges of the fields). Using (\ref{rotated eq}) in (\ref{c in terms of V}) finally leads to 
\beq
c=-2 \cos(2 \theta)\,.
\eeq
To prove (\ref{c General Killing vector }) it is more convenient to redefine the fields so that the Killing vector acts by translations. Note that this is allowed due to the chirality properties of the components of the Killing vector. This, of course, does not preserve the form of the Laplace equation, but instead turns into $\frac{1}{|k^{\phi}|^2}F_{\phi \bar \phi }+\frac{1}{|k^{\chi}|^2}F_{\chi \bar \chi }=0$. Using this in (\ref{c in terms of V}) leads to (\ref{c General Killing vector }). 

\section{Reduction to $\mathcal{N}=(1,1)$}
\label{AppendixN=1}

To reduce to $\mathcal{N}=(1,1)$ (here we follow mostly  \cite{Lindstrom:2007vc, Lindstrom:2008hx,Lindstrom:2005zr}), one decomposes the $\mathcal{N}=(2,2)$ gauge covariant superderivatives into their real and imaginary part, namely
\beq
\nabla_{\pm}=\frac{1}{2}\left(  \mathcal{D}_{\pm} -i Q_{\pm}\right), && \bar{\nabla}_{\pm}=\frac{1}{2}\left(  \mathcal{D}_{\pm} + i Q_{\pm}\right)\,. 
\eeq
Here $\mathcal{D}_{\pm}$ are $\mathcal{N}=(1,1)$ derivatives, which satisfy the algebra 
\beq
\{  \mathcal{D}_{\pm} ,  \mathcal{D}_{\pm} \}= i  \mathcal{D}_{\pm \pm}\,,
\eeq
where $ \mathcal{D}_{\pm \pm}$ is the gauge-covariant space derivative and $Q_{\pm}$ generate the non-manifest supersymmetries. We perform the reduction of the matter fields $\Bbb X_L,\Bbb X_R$ in the covariant approach. That is, we define 
\beq
\Hat{ \Bbb X}_R &=& e^{-\Bbb V_L} e^{i \bar {\tilde{ \Bbb V}}} \Bbb X_R\,, \\  \nonumber
\bar{ \Hat{\Bbb X}}_R &=& \Bbb X_R^{\dagger}e^{- i \tilde{\Bbb V} }\,, \nonumber
\eeq
so that there are no factors $e^{V}$ anywhere. For instance, $ \bar{ \Hat{\Bbb X}}_R \hat{ \Bbb X}_R =  \Bbb X_R^{\dagger}e^{- i \tilde{\Bbb V} }  e^{-\Bbb V_L} e^{i \bar {\tilde{ \Bbb V}}} \Bbb X_R = \bar{\Bbb  X}_R e^{V_R} \Bbb X_R $ and the Lagrangian is simply (dropping the hats) 
\beq
K=\bar{ \Bbb X}_L \Bbb X_L + \bar{ \Bbb X}_R \Bbb X_R +\alpha (\bar{ \Bbb X}_L \Bbb X_R + \bar{ \Bbb X}_R \Bbb X_L)\,.
\eeq
Next, one imposes the fields to be gauge-covariantly semichiral and defines components with gauge-covariant $Q_{\pm}$'s, $i.e.$, 
\beq
X_L = \Bbb X_L\big|\,, \qquad Q_+ \Bbb X_L = i \mathcal{D}_+ \Bbb X_L\,, \qquad Q_- \Bbb X_L \big| = \Psi_{-}\,, \\  
X_R =\Bbb X_R \big|\,, \qquad Q_- \Bbb X_R = i \mathcal{D}_-  \Bbb X_R\,, \qquad Q_+ \Bbb X_R \big| = \Psi_{+}\,.
\eeq
The reduction of the semichiral vector multiplet is given by
\be
d^1=\left( \Bbb F+\bar{\Bbb F} \right) \big|\,, \quad  d^2=\left( \tilde{\Bbb F}+\bar{\tilde{\Bbb F}} \right)\big|\,, \quad   d^3=i \left(  \Bbb F -\bar{\Bbb F} -\tilde{\Bbb F}+\bar{\tilde{\Bbb F}} \right)\big|\,,    \quad   f=-i \left(  \Bbb F -\bar{\Bbb F} +\tilde{\Bbb F}-\bar{\tilde{\Bbb F}} \right)\big|\,,
\ee
from where
\beq
\Bbb F \big|  = \frac{1}{2} \left( d^1+\frac{i}{2} (f-d^3) \right), && \Bbb{\tilde F} \big|  = \frac{1}{2} \left( d^2+\frac{i}{2} (f+d^3) \right)\,.
\eeq
From the definitions  $\Bbb F = i \{\bar \nabla_+, \bar \nabla_-\}$ and $\tilde{\Bbb F } =i \{\bar \nabla_+, \nabla_-\}$, one can solve for the commutation relations
\beq \nonumber
\{Q_+,\mathcal{D}_-\} &=& \mp (d_1+d_2),\, \\ 
\{\mathcal{D}_+,Q_- \} &=& \mp (d_1-d_2)\,, \\  \nonumber
 \{Q_+,Q_- \} &=& \pm d_3\,, \\  \nonumber
 \{\mathcal{D}_+, \mathcal{D}_-\}&=&  f \, ,
\eeq
where the upper(lower) sign is chosen for positive(negative) charge. These are used repeatedly when reducing the matter fields, and the appropriate sign must be chosen depending on the charge of the field it acts on. Note that $f$ is the usual field strength which, in two dimensions, is a total derivative giving the topological charge.

\section{$SU(2)$ symmetry}
\label{AppendixSU2}

As described in the text, the action (\ref{Gauge action for EH}) is invariant under the global $SU(2)$ transformations which rotate $(X^{(1)},  X^{(2)})$ and the cross-ratio (\ref{cross-ratio}) is a natural radial coordinate. At a  fixed radius $R$, we can reach any point by a finite $SU(2)$ transformation from a single point $ X_L^0, X_R^0$. We take $ X_L^0=0$ and $X_R^{0}=\sqrt{R^2-1}$. Thus, by acting with a finite $SU(2)$ transformation, an arbitrary point is parameterized as
\beq \label{parameterization XL,XR}
X_L= \frac{b}{\bar a}\,, && X_R= \frac{a \rho +b}{\bar a - \bar b \rho}\,,
\eeq
where we have defined $\rho^2 \equiv R^2-1$. By means of this identification, the natural remaining invariants are given by the Cartan 1-forms on the group manifold. 
Consider a group element $g$ of $SU(2)$,
\beq \label{SU(2) element}
g=  \left( \begin{array}{cc}  a & b \\ - \bar b & \bar a  \end{array} \right), && |a|^2 + |b|^2 =1\,.
\eeq
The (real) invariant 1-forms $\sigma^{i}$ are defined by
\beq
g^{-1} d g = i  \left( \begin{array}{cc}  \sigma^3 &  \sigma^1+i \sigma^2  \\  \sigma^1- i \sigma^2 & - \sigma^3  \end{array} \right)\,.
\eeq 
In the parameterization (\ref{SU(2) element}), we have
\beq \label{Cartan forms}
\sigma^1=\text{Im} (\bar a d b-  b d \bar a)\,, \,\,\, \sigma^2=-\text{Re} (\bar a d b-  b d \bar a)\,, \,\,\, \sigma^3= - i (\bar a d a +  b d \bar b)\, .
\eeq
The constraint $|a|^2+|b|^2=1$ ensures the reality of $\sigma^3$. From (\ref{parameterization XL,XR}) and (\ref{Cartan forms})   we find
\beq  \nonumber
dX_L &=& \frac{1}{ \bar a ^2}(i \sigma^1 - \sigma^2)\,, \\
dX_R&=&\frac{1}{(\bar a - \rho \bar b)^2}\left[2 i \rho \sigma^3 + i (1-\rho^2) \sigma^1 - (1+\rho^2) \sigma^2+ d \rho \right]\, . 
\eeq
These are the expressions which allow us to rewrite the Eguchi-Hanson metric in an explicitly $SU(2)$-invariant form. Another well-known property of Eguchi-Hanson is that its complex structures are preserved by the $SU(2)$ (in the Taub-NUT case they form a triplet). The Lie derivative along  $\xi$ of a $(1,1)$ tensor such as a complex structure is given by
\beq
\mathcal{L}_{\xi} J_{\pm} = \xi J_{\pm} - [\partial \cdot \xi ,J_{\pm}] \,, && \partial \cdot \xi \equiv \left( \begin{array}{cc} \partial_L  \xi^L &0 \\ 0 &  \partial_R  \xi^R \end{array} \right)\,, \label{Lie derivative}
\eeq
where 
\beq
 \partial_L  \xi^L \equiv \left( \begin{array}{cc}  \partial_l \xi^l &0 \\ 0 &   \partial_{\bar l} \bar \xi^l \end{array} \right), && \partial_R  \xi^R \equiv \left( \begin{array}{cc}  \partial_r \xi^r &0 \\ 0 &   \partial_{\bar r} \bar \xi^r \end{array} \right)\,.
\eeq
The equations from $\mathcal{L}_{\xi} J_{+}=0$ read
\beq  \nonumber
\xi^{\mu}  \partial_{\mu}( \mathcal{K}_{RL}^{-1} C_{LL})  - ( \partial_R  \xi^R  \mathcal{K}_{RL}^{-1} C_{LL}  -  \mathcal{K}_{RL}^{-1} C_{LL}   \partial_L  \xi^L)&=&0\,, \\[2mm]
\xi^{\mu}  \partial_{\mu}(\mathcal{K}_{RL}^{-1}J_s \mathcal{K}_{LR})  - [ \partial_R  \xi^R,  \mathcal{K}_{RL}^{-1}J_s \mathcal{K}_{LR} ]&=&0\,,
\eeq
and similarly for $J_-$, exchanging $R$ by $L$.  We verified that these equations are satisfied by explicit calculations from the potential (\ref{KEHp}).

\end{document}